\documentclass[
twocolumn]{aastex701}

\usepackage{amsmath, amssymb}

\begin{document}

\title{The Mysterious Inspiral of WASP-12\,b: Why Obliquity Tides Cannot Drive Orbital Decay}

\author[0000-0001-9985-0643]{Caleb Lammers}
\affiliation{Department of Astrophysical Sciences, Princeton University, 4 Ivy Lane, Princeton, NJ 08544, USA}
\email{caleb.lammers@princeton.edu}

\author[0000-0001-8283-3425]{Yubo Su}
\affiliation{Department of Astrophysical Sciences, Princeton University, 4 Ivy Lane, Princeton, NJ 08544, USA}
\affiliation{Canadian Institute for Theoretical Astrophysics, University of Toronto, 60 St George Street, Toronto, M5S 3H8 Ontario, Canada}
\email{yubo.su@utoronto.ca}

\author[0000-0002-4265-047X]{Joshua N.\ Winn}
\affiliation{Department of Astrophysical Sciences, Princeton University, 4 Ivy Lane, Princeton, NJ 08544, USA}
\email{jnwinn@princeton.edu}

\shorttitle{WASP-12\,b: Not Obliquity Tides}
\shortauthors{Lammers, Su, \& Winn}

\begin{abstract}

WASP-12\,b's orbit is decaying, for unknown reasons. The planet's period is shrinking more rapidly than can be attributed to equilibrium tides or dynamical tides in a main-sequence star. Planetary obliquity tides could be sufficiently dissipative to drive WASP-12\,b's inspiral, but would also damp the planet's obliquity, halting the decay. Millholland \& Laughlin proposed that a nearby, low-mass planet (${\sim}\,10$\,M$_\oplus$) is maintaining a large obliquity for WASP-12\,b, sustaining the dissipation. We re-evaluated this hypothesis, finding that the companion must be more massive than originally proposed (${\gtrsim}\,65$\,M$_\oplus$) to absorb WASP-12\,b's orbital angular momentum. Radial velocity data allowed us to rule out a companion of this type. Any companions within $3$~AU have $K\,{\lesssim}\,14$\,m/s at $95$\% confidence.

\end{abstract}

\keywords{\uat{exoplanets}{498} --- \uat{hot Jupiters}{753} --- \uat{star-planet interactions}{21} --- \uat{tides}{1702}}

\section{Introduction}
\label{sec:intro}

Hot Jupiters orbit so close to their host stars that tidal interactions should shrink their orbits \citep{Rasio1996, Sasselov2003, Levrard2009}. In principle, inspiral can be studied directly by monitoring the periods of transiting hot Jupiters for long-term changes. Orbital decay has been securely detected for only one planet: WASP-12\,b \citep{Maciejewski2016, Patra2017, Yee2020}. The system consists of a late-F star \citep{Hebb2009} that hosts a hot Jupiter with an orbital period $1.09$~days, mass $1.47~\mathrm{M_J}$, and radius $1.90~\mathrm{R_J}$ \citep{Collins2017}. Its orbital period is steadily shrinking at a rate of $\dot{P}\,{=}\,-29\,{\pm}\,2$\,ms\,yr$^{-1}$, limiting the planet's remaining lifespan to ${\lesssim}\,3$\,Myr.

At first glance, WASP-12\,b's inspiral seems to confirm a long-standing theoretical prediction. However, there is a significant mismatch between tidal theory and the measured decay rate. The observed inspiral is several orders of magnitude faster than can be explained by equilibrium stellar tides \citep{Bailey&Goodman2019}, eccentricity tides \citep{Maciejewski2020}, or tidal resonance locking \citep{Ma&Fuller2021}. If WASP-12 were a subgiant, dynamical tides could explain the planet's rapid inspiral \citep{Weinberg2017, Barker2020}, but stellar modeling has consistently indicated that WASP-12 is on the main sequence (\citealt{Bailey&Goodman2019, Efroimsky&Makarov2022, Leonardi2024, Golonka2026}; P.\ McCreery \& S.\ P.\ Schmidt et al.\ submitted; E.\ Y.\ Zhang et al.\ in prep).

\citet{Millholland&Laughlin2018} proposed an alternative mechanism in which planetary obliquity tides provide the source of dissipation (see Section~\ref{sec:obliquity_tides}). This idea has its roots in Solar System dynamics. The Moon's 7$^\circ$ obliquity is maintained by a resonance between the precession of its spin axis and orbital plane \citep{Colombo1966}, and Saturn's $27^\circ$ obliquity is thought to have been excited by a related resonance involving Neptune's nodal precession \citep{Ward&Hamilton2004, Hamilton&Ward2004}. Such spin-orbit resonances are known as ``Cassini states.'' In the first application of Cassini states to exoplanets, \citet{Winn&Holman2005} proposed that obliquity tides might provide enough internal heating to explain the anomalously large radius of the hot Jupiter HD\,209458\,b.

\citet{Fabrycky2007} found a fatal flaw in the obliquity-tide hypothesis for hot Jupiter inflation. The heat is ultimately derived from orbital energy, and extracting enough energy to inflate the planet would also transfer orbital angular momentum to the body that maintains orbital precession, driving the system out of the high-obliquity Cassini state (see also \citealt{Levrard2007, Peale2008}). In \citet{Millholland&Laughlin2018}'s revival of the Cassini-state idea, they proposed a companion planet to circumvent the issues with the hot Jupiter inflation theory. Here, we show that a problem related to angular momentum also plagues the obliquity-tide hypothesis for orbital decay.

\section{The obliquity-tide hypothesis}
\label{sec:obliquity_tides}

\begin{figure*}
\centering
\includegraphics[width=0.95\textwidth]{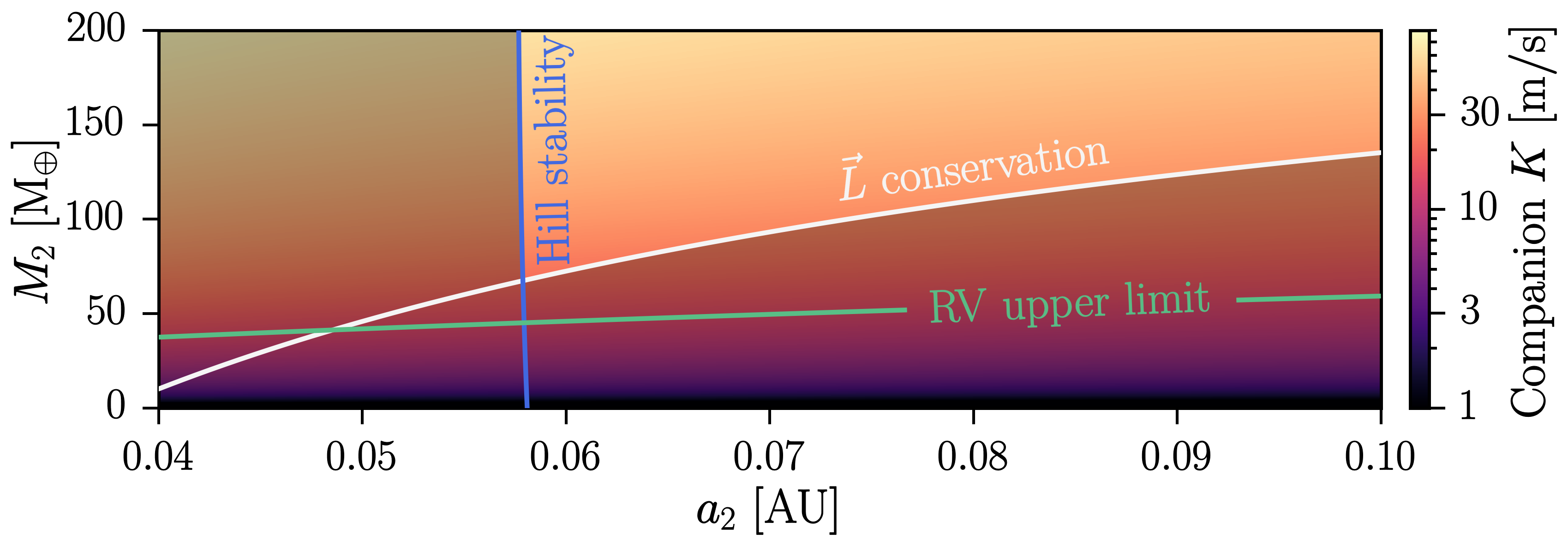}
\caption{Constraints on the hypothetical planet in the obliquity-tide scenario, under assumptions that lend the hypothesis maximum credence. RV amplitudes are indicated by the background color (assuming $i_2\,{=}\,90^\circ$). The blue line shows the minimum $a_2$ compatible with Hill stability, and the white line is the lower limit imposed by angular-momentum conservation, leaving the brightly colored region as the parameter space allowed by the theory. The green line is the upper limit imposed by the RV data, which excludes all of the allowed parameter space.}
\label{fig:perturber_params}
\end{figure*}

In the absence of companions, tidal dissipation would drive WASP-12\,b toward synchronous rotation ($\omega_1\,{\approx}\,n_1$), zero obliquity ($\theta_1\,{\approx}\,0^\circ$), and a circular orbit ($e_1\,{\approx}\,0$), after which planetary tidal dissipation would cease. \citet{Millholland&Laughlin2018} proposed that an inclined exterior companion changes this outcome. The companion forces WASP-12\,b's orbital angular momentum $\vec{L}_1$ to precess around the total angular-momentum vector $\vec{L}_{\rm tot}$
at the approximate rate
\begin{equation}
    \label{eqn:nodal_prec}
    g_1 \approx -\frac{1}{4}b_{3/2}^{(1)}(\alpha_{12}) \, \alpha_{12} \left(n_1 \alpha_{12} \frac{M_2}{M_\star} + n_2 \frac{M_1}{M_\star} \right),
\end{equation}
where $\alpha_{12}\,{=}\,a_{1}/a_2$, $b_{3/2}^{(1)}(\cdot)$ is the standard Laplace coefficient, the subscript $2$ refers to the companion, and the minus sign indicates nodal regression. Meanwhile, the star torques WASP-12\,b's rotational bulge, causing the planet's spin axis to precess around $\vec{L}_{1}$ at a rate $\alpha_1 \cos \theta_1$, where
\begin{equation}
    \label{eqn:spin_prec}
    \alpha_1 \approx \frac{1}{2}\frac{M_\star}{M_1}\left(\frac{R_1}{a_1}\right)^3 \frac{k_{2}}{\mathcal{C}} \, \omega_1.
\end{equation}
Here, $k_2$ is WASP-12\,b's Love number and $\mathcal{C}$ is its normalized moment of inertia. We adopt the same fiducial values as \citet{Millholland&Laughlin2018}, $k_2\,{\approx}\,0.1$ and $\mathcal{C}\,{\approx}\,0.2$.\footnote{For reference, Jupiter's Love number is $k_2\,{\approx}\,0.57$ \citep{Durante2020}. Adopting this value would make it even harder to satisfy the constraints on the companion presented in Section~\ref{sec:ang_mom_problem}.}

As weak equilibrium tides slowly shrink the orbit of WASP-12\,b, the two precession rates evolve in opposite directions: $|g_1|$ decreases and $\alpha_1$ increases. Eventually, the system slowly passes through the condition
\begin{equation}
    \label{eqn:resonance_condition}
    \frac{|g_1|}{\alpha_1} \approx \cos \theta_1,
\end{equation}
capturing the planet in a spin-orbit resonance. As $a_1$ and $|g_1|/\alpha_1$ continue to fall, Equation~\ref{eqn:resonance_condition} requires $\cos\theta_1$ to decrease. That is, the spin axis must adjust by tipping toward higher obliquity. This introduces a new dissipation mechanism, obliquity tides, which can drive rapid inspiral:
\begin{align}
    \label{eqn:inspiral_rate}
\frac{\dot{a}}{a} = &-\frac{1}{5\,\mathrm{Myr}} \left(\frac{Q_p'}{5\,{\times}\,10^6}\right)^{-1} \left(\frac{M_\star}{1.43\,\mathrm{M_\odot}}\right)^{3/2} \left(\frac{M_1}{1.5\,\mathrm{M_J}}\right)^{-1} \nonumber \\ &\left(\frac{R_1}{1.9\,\mathrm{R_J}}\right)^{5} \left(\frac{a_1}{0.023\,\mathrm{AU}}\right)^{-13/2} \left(\frac{\sin^2 \theta_1}{1 + \cos^2 \theta_1}\right),
\end{align}
where $Q_p'\,{=}\,3Q_p/(2k_2)$ is the planet's modified tidal quality factor. The crucial assumptions are that this resonance was encountered after ${\gtrsim}\,1$\,Gyr, allowing the rapid-inspiral phase to be observable today, and that the system thereafter evolved adiabatically, remaining in resonance. For more details, see \citet{Millholland&Laughlin2018} and \citet{Su&Lai2022}.

\section{The angular-momentum problem}
\label{sec:ang_mom_problem}

As WASP-12\,b's orbital radius shrinks from $a_{1,\mathrm{init}}$ to $a_{1,\mathrm{now}}\,{=}\,0.023\,\mathrm{AU}$, the change in orbital angular momentum must be compensated elsewhere in the system. Stellar spin-up is not an appealing possibility because it requires a seemingly implausible stellar dissipation rate, the problem that obliquity tides are meant to solve (see also Section~\ref{sec:ang_mom_discussion}). \citet{Millholland&Laughlin2018} proposed balancing the angular-momentum budget by aligning the orbital angular momenta of the hot Jupiter and its companion ($\vec{L}_1$ and $\vec{L}_2$) and possibly also the star's spin angular momentum ($\vec{S}_\star$). Below, we examine the angular-momentum ledger, making assumptions as favorable as possible for the obliquity-tide hypothesis.

To maximize the angular momentum that can be absorbed by reorienting $\vec{L}_1$, $\vec{L}_2$, and $\vec{S}_\star$, we assume $\vec{L}_1$, $\vec{L}_2$, and $\vec{S}_\star$ were initially mutually perpendicular and that $\vec{L}_1$ and $\vec{L}_2$ are now closely aligned (though not perfectly aligned, or else the spin-orbit resonance could not be maintained).\footnote{A retrograde initial configuration might allow reorientation with $\vec{L}_1$ to absorb even more angular momentum. However, secular torques would drive a retrograde configuration towards antialignment, rather than alignment.} Denoting by $\theta_\star$ the angle between $\vec{L}_1$ and $\vec{S}_\star$, conservation of angular momentum requires
\begin{align}
    &a_{1,\mathrm{init}} < a_{1,\mathrm{now}} + \frac{2M_2}{M_1}\sqrt{a_{1,\mathrm{now}}\,a_2}~+ \nonumber \\
    &\frac{2 S_\star \cos \theta_\star} {M_1}\sqrt{\frac{a_{1,\mathrm{now}}}{G M_\star}} + \frac{2M_2 S_\star\cos\theta_\star}{M_1^2}
        \sqrt{\frac{a_2}{GM_\star}}.\label{eqn:ang_mom_cons} 
\end{align}
Since WASP-12\,b is transiting, $\vec{L}_1$ lies close to the sky plane, and the projection of $\vec{S}_\star$ in that plane is
\begin{equation}
    \label{eqn:S_star}
    S_\star \cos \theta_\star \approx \mathcal{C}_\star M_\star R_\star (v \sin i_\star) \cos \lambda,
\end{equation}
where $\mathcal{C}_\star\,{\approx}\,0.07$ is the star's normalized moment of inertia and $v \sin i_\star$ and $\lambda$ are its projected rotation velocity and obliquity. To lend the obliquity-tide hypothesis maximum credence, we assume $\cos \lambda\,{\approx}\,1$ and $v \sin i_\star\,{\approx}\,1.9$\,km/s, the upper limit from \citet{Leonardi2024}.

Given the system's finite capacity to absorb WASP-12\,b's orbital angular momentum, we now consider the competing requirements on the hypothetical companion. A massive, distant companion would enlarge the angular-momentum reservoir, but its semimajor axis also determines where WASP-12\,b encountered the spin-orbit resonance, and therefore how much angular momentum it has lost since then. Placing the companion too far away therefore risks exceeding the angular-momentum budget. Yet, placing the companion too close to WASP-12\,b risks dynamical instability. We evaluated these criteria on a 2D grid of ($a_2$, $M_2$) values, using the spin-orbit resonance condition (Equation~\ref{eqn:resonance_condition}) to determine WASP-12\,b's initial semimajor axis $a_{1,\mathrm{init}}$ and enforce the angular-momentum constraint (Equation~\ref{eqn:ang_mom_cons}). We also impose the Hill-stability criterion of \citet{Gladman1993}, appropriate for two nearly circular, nearly coplanar planets. This criterion should be regarded as a lower bound on the required separation; the mutually inclined configurations probably require wider spacings for long-term stability than coplanar configurations \citep[e.g.,][]{VerasArmitage2004}. Figure~\ref{fig:perturber_params} shows that to satisfy both constraints, the companion must have $a_2\,{\gtrsim}\,0.06$\,AU and $M_2\,{\gtrsim}\,65$\,M$_\oplus$. For comparison, \cite{Millholland&Laughlin2018} predicted $a_2\,{\approx}\,0.04$\,AU and $M_2\,{\approx}\,10$\,--\,$20\,\mathrm{M_\oplus}$ based on resonance and dissipation requirements, but without considering angular-momentum or stability constraints. The angular-momentum constraint increases the required mass of the companion and makes it more readily detectable, motivating a close inspection of WASP-12's radial velocity (RV) data.

\begin{figure*}
\centering
\includegraphics[width=0.95\textwidth]{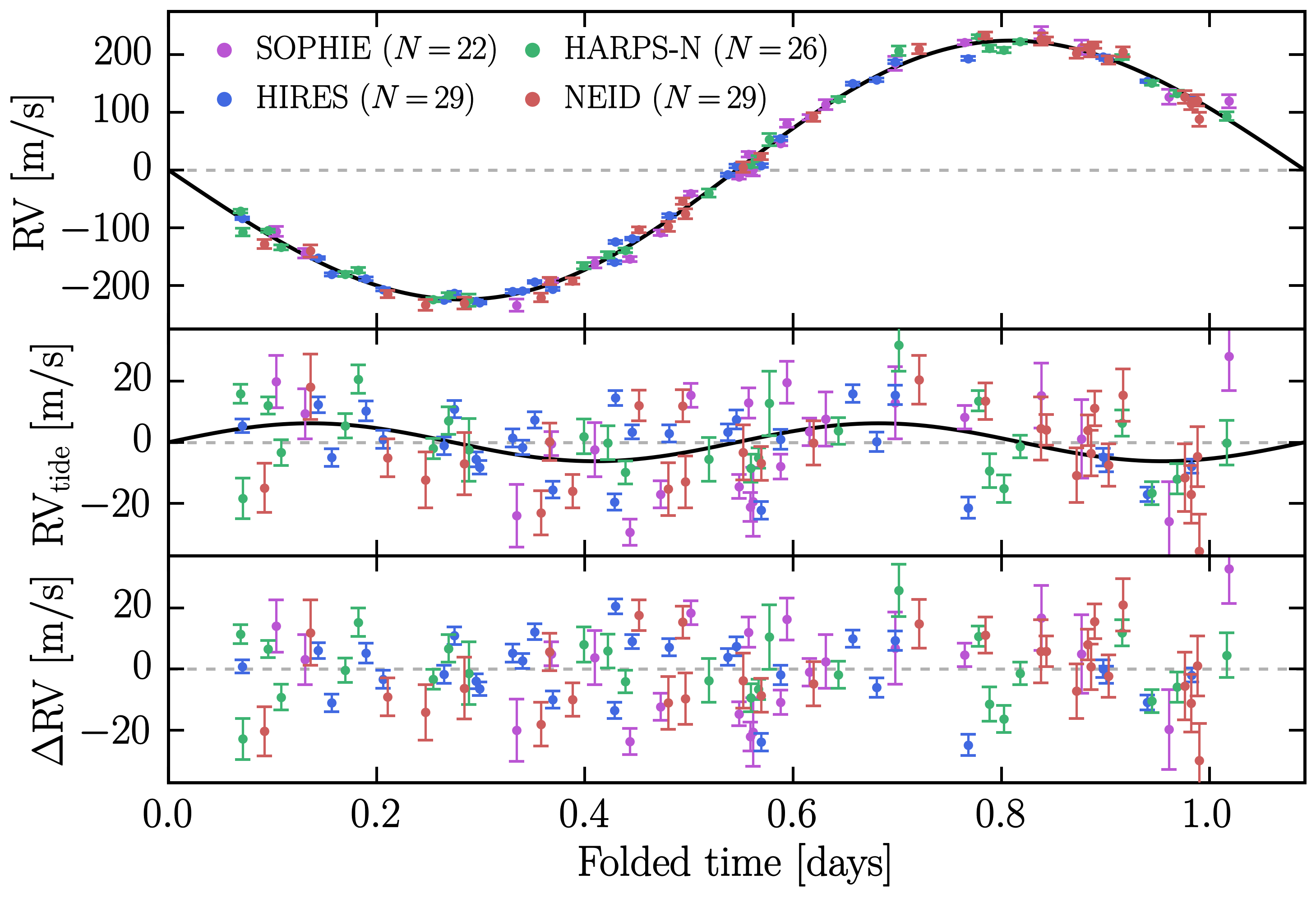}
\caption{{\it First panel:} WASP-12's RV dataset, folded with respect to the period of WASP-12\,b (transits occur at $t\,{=}\,0$). The best-fitting ``WASP-12\,b + tidal bulge'' model is plotted in black. {\it Second panel:} Tidal RV signal, obtained by subtracting the RV model with $K_\mathrm{tide}\,{=}\,0$. The $6$~m/s modulation due to the star's tidal bulge is detected, illustrating the sensitivity of the dataset to low-amplitude signals. {\it Third panel:} RV residuals. All error bars show formal uncertainties, before inflation with jitter terms (Table~\ref{table:RV_parameters}). The residuals appear randomly distributed around zero, with underestimated uncertainties but no obvious trends.}
\label{fig:folded_RVs}
\end{figure*}

\section{RV analysis}
\label{sec:RVs}

We compiled RV measurements from four instruments: SOPHIE data from \citet{Hebb2009} and \citet{Husnoo2011}, HARPS-N data from \citet{Bonomo2017} and \citet{Maciejewski2020}, HIRES data from \citet{Yee2020}, and NEID data from \citet{WinnStefansson2025}, supplemented
by four new RVs (Appendix~\ref{sec:RV_dataset}). The NEID spectra were processed with version 1.5.2 of the data reduction pipeline.\footnote{\url{https://neid.ipac.caltech.edu/docs/NEID-DRP/}} RVs obtained during transits and those with unusually large formal uncertainties were discarded, as was a single SOPHIE RV that was a large outlier. Whenever multiple RVs were obtained in a single night, only the data point with the lowest formal uncertainty was retained. The resulting 106 RVs are shown in Figure~\ref{fig:folded_RVs} and given in Table~\ref{table:RV_data}.

\subsection{WASP-12\,b}
\label{sec:WASP12b}

We fit the RV data with a two-sinusoid model: 
\begin{equation}
    \label{eqn:1pl_RV_model}
    \mathrm{RV_{1pl}}(t) = -K_1 \sin \phi(t) + K_\mathrm{tide} \sin 2\phi(t) + \gamma_i,
\end{equation}
where the first term is WASP-12\,b's orbital signal (assuming $e_1\,{=}\,0$), the second term is the signal induced by the star's tidal bulge \citep{Arras2012, Maciejewski2020}, and $\gamma_i$ is the instrument-specific RV offset. Here, $\phi$ represents WASP-12\,b's orbital phase, calculated according to a linear-decay ephemeris
\begin{equation}
    \label{eqn:WASP12b_orbital_phase}
    \phi(t) = \frac{4\pi(t - T_0)}{P_1 + \sqrt{P_1^2 + 2 \frac{dP_1}{dN}(t - T_0)}}.
\end{equation}
The decay rate $dP_1/dN$ and the period $P_1$ at reference time $T_0$ were taken from \citet{Wong2022}. We optimized the amplitudes of the two sinusoids ($K_1$ and $K_\mathrm{tide}$), along with a zero-point and excess noise (``jitter'') term for each of the four instruments. Fitting was performed using the Nelder-Mead optimizer from \texttt{scipy.optimize.minimize} \citep{Virtanen2020}, and uncertainties were calculated using an affine-invariant Markov Chain Monte Carlo (MCMC) analysis with the \texttt{emcee} code \citep{Foreman-Mackey2013}. We used $200$ walkers, which each took $20{,}000$ steps (${>}\,100$ autocorrelation lengths), the first $5{,}000$ of which were discarded.

The best-fit RV model is plotted in Figure~\ref{fig:folded_RVs}, and the MCMC constraints are summarized in Table~\ref{table:RV_parameters}. This model assigns each RV dataset an excess noise of about $10$~m/s, which reduces the model's $\chi^2$ from $725$ to $102$, in concordance with the $106\,{-}\,10$ degrees of freedom. $K_1$ is tightly constrained, and the tidal bulge signal is detected with a median $K_\mathrm{tide}$ that differs from zero by $3.7\sigma$.

\subsection{Search for companions}
\label{sec:companion_search}

\begin{deluxetable}{cc}
\tablecaption{RV Model Fit Parameters\label{table:RV_parameters}}
\tablehead{\colhead{Parameter} & \colhead{Value}}
\startdata
Orbital RV amplitude $K_1$ & $224.1^{+1.7}_{-1.7}$~m\,s$^{-1}$ \\
Tidal RV amplitude $K_\mathrm{tide}$ & $6.2^{+1.7}_{-1.7}$~m\,s$^{-1}$ \\
\tablenotemark{a}Planet mass $M_1$ & $1.454^{+0.068}_{-0.069}~\mathrm{M_J}$ \\
\hline
Excess RV noise $\sigma_\mathrm{{SOPHIE}}$ & $14.1^{+3.3}_{-2.6}$~m\,s$^{-1}$ \\
Excess RV noise $\sigma_\mathrm{{HARPSN}}$ & $9.4^{+2.1}_{-1.7}$~m\,s$^{-1}$ \\
Excess RV noise $\sigma_\mathrm{{HIRES}}$ & $10.4^{+1.8}_{-1.4}$~m\,s$^{-1}$ \\
Excess RV noise $\sigma_\mathrm{{NEID}}$ & $10.1^{+2.4}_{-2.0}$~m\,s$^{-1}$ \\
\hline
\tablenotemark{b}Orbital period $P_1$ & $1.091419370$~days \\
\tablenotemark{b}Midtransit time & $2457103.283654$~BJD \\
\tablenotemark{b}Decay rate $dP_1/dN$ & $-1.031\,{\times}\,10^{-9}~\mathrm{days/orbit}$ \\
Orbital eccentricity $e_1$ & $0.00$
\enddata
\tablenotetext{a}{Assuming $M_\star\,{=}\,1.434\,{\pm}\,0.1\,M_\odot$ and $i_1\,{=}\,83.37\,{\pm}\,0.68^\circ$ \citep{Collins2017}.}
\tablenotetext{b}{Fixed parameters, taken from \citet{Wong2022}.}
\end{deluxetable}

We searched for companion planets by fitting the RVs with a three-component model:
\begin{equation}
    \label{eqn:RV_model}
    \mathrm{RV_{2pl}}(t) = \mathrm{RV_{1pl}}(t) + c_1 \sin \frac{2\pi t}{P_2} + c_2 \cos \frac{2\pi t}{P_2},
\end{equation}
where $c_1\,{=}\,K_2 \cos \phi_2$ and $c_2\,{=}\,K_2 \sin \phi_2$. This model assumes that the companion's orbit is nearly circular. For a first pass, we held the RV jitter terms fixed at the best-fit values from Section~\ref{sec:WASP12b}, which makes the optimization problem linear. We scanned across approximately $26{,}000$ possible orbital periods for the companion, which span semimajor axes from $0.03$ to $30$~AU. For each period, we performed weighted least-squares optimization of the eight free parameters ($K_1$, $K_\mathrm{tide}$, $c_1$, $c_2$, and four RV zero-points), and we recorded the model's Bayesian information criterion (BIC) statistic.\footnote{$\mathrm{BIC}\,{=}\,k \ln(n)\,{-}\,2\ln(\mathcal{L})$, where $k$ is the number of free parameters, $n$ is the number of data points, and $\ln(\mathcal{L})$ is the log-likelihood \citep{Schwarz1978}.} Fixing the jitter terms prevents the companion model from absorbing excess noise, which systematically raises the BIC values. For the five lowest-BIC companion periods, we performed non-linear optimization (with Nelder-Mead) of the four jitter terms, linearly optimizing the other eight parameters at each optimizer step. The speedup from this two-step procedure was necessary for the inject-and-recover experiments described below. We obtained similar results, at a much greater computational cost, when we performed non-linear optimization at each of the $26{,}000$ possible periods.

The search did not uncover evidence for companions. The largest $\Delta \mathrm{BIC}$ improvement with respect to the single-planet model was $17.2$. This was only narrowly larger than the next largest $\Delta \mathrm{BIC}$ peaks; a total of $13$ different periods had $\Delta \mathrm{BIC}\,{>}\,10$. We determined the appropriate $\Delta \mathrm{BIC}$ threshold by extrapolating from the distribution of $\Delta \mathrm{BIC}$ peaks, as described by \citet{Howard&Fulton2016}. The power-law fit to the $\Delta \mathrm{BIC}$ histogram implied a $1$\% chance of obtaining a $\Delta \mathrm{BIC}\,{=}\,25$ peak by chance, which we adopted as our nominal detection threshold.

\begin{figure}
\centering
\includegraphics[width=0.475\textwidth]{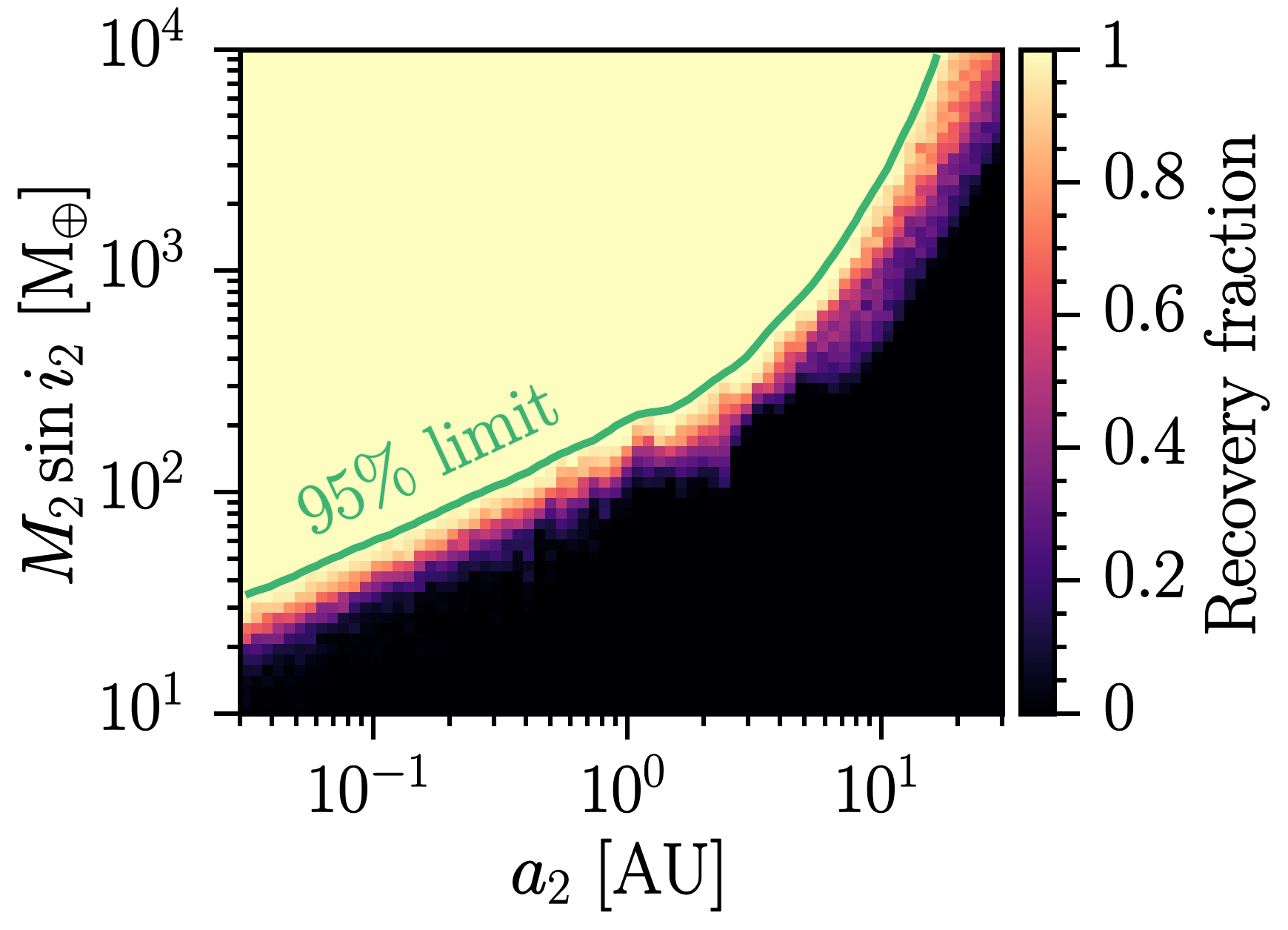}
\caption{RV constraints on companions of WASP-12\,b. The color conveys the fraction of injected signals that were recovered with $\Delta \mathrm{BIC}\,{>}\,25$. The green contour traces the $95$\% recovery fraction limit. Companions with $a_2\,{\approx}\,0.06$\,AU must have $M_2 \sin i_2\,{\lesssim}\,46\,\mathrm{M_\oplus}$ (i.e., $K\,{\lesssim}\,14$\,m/s). Separately, any companions with $M_2 \sin i_2\,{\gtrsim}\,13\,\mathrm{M_J}$ must be located beyond $10$~AU.}
\label{fig:BIC_periodogram}
\end{figure}

Finally, we performed inject-and-recover experiments to determine the constraints placed by the RV data on companions to WASP-12\,b. We sampled the companion's $M_2 \sin i_2$ log-uniformly from ($10$,\,$10^4\,\mathrm{M_\oplus}$) and $a_2$ log-uniformly from ($0.03$,\,$30\,\mathrm{AU}$). Then, we added the companion's signal to the RV data and performed the BIC-periodogram search for companions, as described above. We repeated this procedure for about $3\,{\times}\,10^6$ trials. Figure~\ref{fig:BIC_periodogram} shows the fraction of trials in which companions of different types produced a peak in the periodogram exceeding $\Delta \mathrm{BIC}\,{=}\,25$. For $a_2\,{\approx}\,0.06$\,AU, the orbital radius at which the obliquity-tide scenario can be sustained by the lowest-$M_2$ companion (Figure~\ref{fig:perturber_params}), the RV data rule out companions with $M_2 \sin i_2\,{\gtrsim}\,46\,\mathrm{M_\oplus}$ at $95$\% confidence. Thus, the parameter space for companions that meet the criteria required by the obliquity-tide hypothesis is excluded.

A low line-of-sight inclination could allow the companion's true mass to exceed the RV upper limit on $M_2\sin i_2$. However, this cannot easily rescue the obliquity-tide hypothesis. Because WASP-12\,b transits, lowering $i_2$ tends to increase the present-day mutual inclination between the two planetary orbits, thereby reducing the amount of angular momentum that could have been absorbed by reorienting $\vec{L}_2$ toward $\vec{L}_1$. A companion that satisfies the RV constraints and the constraints from Section~\ref{sec:ang_mom_problem} requires an unlikely three-dimensional geometry: a near-perpendicular orbit for which ${\lesssim}\,10$\% of azimuthal orientations would evade detection. Less favorable assumptions about dynamical stability, $k_2$, or the initial orientations of $\vec{L}_2$ and $\vec{S}_\star$ would make this scenario even less likely.

Independent of WASP-12\,b's orbital decay, constraints on massive far-out companions are of interest because such companions are required in some theories for the formation of hot Jupiters \citep[e.g.,][]{Knutson2014, Bryan2016}. Any companions of WASP-12\,b within $3$~AU must have $K\,{\lesssim}\,14$\,m/s. The long baseline of the RV dataset rules out edge-on brown dwarf companions out to about $10$~AU. It is worth noting that $e_2\,{\approx}\,0$ is a reasonable assumption for small, close-in planets \citep{VanEylen&Albrecht2015, Hadden&Lithwick2017}, but it is suspect for cold giant companions \citep{Butler2006, Kipping2013}. Relaxing this assumption would weaken the constraints, at least slightly.

\section{Discussion}
\label{sec:discussion}

\subsection{Other angular-momentum reservoirs}
\label{sec:ang_mom_discussion}

Section~\ref{sec:ang_mom_problem} considered the most favorable possible reorientation of $\vec{L}_2$ and $\vec{S}_\star$. There are a few other conceivable reservoirs for the angular momentum lost by WASP-12\,b's orbit, but none can save the theory:

\begin{itemize}

\item Spinning up the star. As noted earlier, the obliquity-tide hypothesis was proposed because WASP-12\,b's decay defies the expectations for dissipation within the star. The ratio of the strength of planetary obliquity tides to stellar tides is ${\sim}\,500$ \citep{Millholland&Laughlin2018}, suggesting that only ${\sim}\,0.2$\% of the angular-momentum exchange occurs through stellar spin-up instead of reorientation. See also footnote~1 of \citet{Fabrycky2007} for a general argument for why planetary tides cannot substantially alter $S_\star$.

\item Spinning up the companion. This can be rejected because the upper limit on the companion's mass prevents it from bearing enough angular momentum ($S_2\,{\sim}\,10^{-3}S_\star$).

\item Widening the companion's orbit. Here, the difficulty is that the secular planet-planet torques invoked to maintain the spin-orbit resonance do not alter the planets' semimajor axes. Non-secular interactions would be required. One possibility is the crossing of mean-motion resonances (MMRs) during WASP-12\,b's inspiral. A resonance crossing can give the companion a small outward kick, increasing $a_2$ and $L_2$. To first order,
\begin{equation}
    \label{eqn:a2_kick_MMRs}
    \frac{\Delta L_2}{L_2} \approx \frac{2^{1/3}}{2} \frac{8 r^{2/3}}{3^{2/3}} \left(\frac{M_1}{M_\star}\right)^{2/3} (j + 1)^{1/3},
\end{equation}
for a $j\,{+}\,1$:$j$ MMR (based on Equation~60 from \citealt{Petit2017}), where $r\,{\approx}\,0.8$ is a numerical constant. The $3$:$2$ MMR could be crossed in the obliquity-tide scenario, but Equation~\ref{eqn:a2_kick_MMRs} predicts only a $3$\% rise in $L_2$, far too small to balance the angular-momentum budget. Furthermore, Equation~\ref{eqn:a2_kick_MMRs} probably overestimates $\Delta L_2$ because it assumes that $a_2$ is increased by the maximum possible amount (the full resonance width).

\item Circularizing the companion's orbit. This reservoir is also too small: the close-in companion must have $e_2\,{\lesssim}\,0.3$ to avoid crossing WASP-12\,b's pre-inspiral orbit. Decreasing $e_2$ from $0.3$ to zero only increases $L_2$ by $5$\%.

\item Mass lost by WASP-12\,b \citep[e.g.,][]{Fossati2010, Haswell2012, Nichols2015, Bell2019}. Escaping gas could reduce the planet's orbital angular momentum. Whether this angular momentum is permanently lost from the orbit, or partly returned through subsequent interactions with the planet, is uncertain \citep{Weldon2026, HallattMillholland2026}. In any case, mass loss is only expected to be relevant near WASP-12\,b's present orbit, where the planet is strongly irradiated and close to Roche-lobe overflow. It does not provide an obvious sink for the angular momentum lost earlier in the inspiral.

\end{itemize}

\subsection{Other capture scenarios}
\label{sec:chance_capture_CS2}

The calculation in Section~\ref{sec:ang_mom_problem} assumed that WASP-12\,b encountered the spin-orbit resonance in a synchronous state, with $\theta_1\,{\approx}\,0^\circ$ and $\omega_1\,{\approx}\,n_1$. In principle, WASP-12\,b could have entered the resonance before its spin was synchronized. Although rapid rotation would worsen the angular-momentum problem, a large initial obliquity would help by lowering $\cos \theta_1$ and allowing a more distant or less massive companion to produce a matching value of $g_1$. However, the high obliquity cannot be primordial --- if WASP-12\,b encountered the spin-orbit resonance shortly after forming, the ${\lesssim}\,100$\,Myr required to reach its current orbital distance conflicts with WASP-12's multi-Gyr age.\footnote{It follows from Equation~\ref{eqn:inspiral_rate} that the time required to inspiral from $a_{1,\mathrm{init}}$ to $a_{1,\mathrm{now}}$ is $\Delta t\,{=}\,(2/13)(a_{1,\mathrm{now}}/\dot{a}_{1,\mathrm{now}})\left[1 - (a_{1,\mathrm{init}}/a_{1,\mathrm{now}})^{13/2}\right]$. Substituting measured values and applying angular momentum and RV constraints yields $\Delta t\,{\lesssim}\,100$\,Myr.} Thus, we are left with a ``chicken-and-egg'' problem: today's large obliquity requires WASP-12\,b to have obtained a large obliquity before it entered the resonance but ${\gtrsim}\,1$\,Gyr after forming.

\subsection{Other systems}
\label{sec:other_systems}

We have argued against obliquity tides as the explanation for WASP-12\,b's orbital decay. Could this mechanism operate in other systems, as some authors have suggested \citep[e.g.,][]{Vissapragada2022, Hagey2025}? A more rapidly rotating star would help by providing a larger angular-momentum reservoir and allowing a lower-mass companion to balance the angular-momentum budget. Still, there are significant challenges, including that the hot Jupiter must be accompanied by an exterior companion and begin on an orbit that is misaligned with the stellar equator. No such systems are known, and nearby outer companions to hot Jupiters seem rare \citep{Steffen2012, Hord2021, Sha2026}.

Companions of arbitrarily low mass cannot maintain the spin-orbit resonance. Tidal dissipation displaces the planet's spin axis by an angle $\psi$ out of the plane containing $\vec{L}_1$ and $\vec{L}_\mathrm{tot}$. Once $\psi\,{>}\,90^\circ$, the companion's precessional torque can no longer balance the tidal alignment torque \citep{Fabrycky2007, Levrard2007}. For $L_2\,{\ll}\,L_1$, the resonance is only maintained when
\begin{align}
\label{eqn:resonance_stability}
    M_2 \gtrsim &\frac{3}{Q_p'} \frac{M_\star^3}{M_1^2} \left(\frac{R_1}{a_1}\right)^{9/2} \left(\frac{a_2}{a_1}\right)^{19/4} \nonumber \\
    &\frac{(k_2/3)^{1/2}}{[\mathcal{C} f(\alpha_{12})]^{3/2} \sin i_\mathrm{mut} \cos^{3/2} i_\mathrm{mut}},
\end{align}
where $f(\alpha_{12})\,{=}\,b^{(1)}_{3/2}(\alpha_{12})/(3 \alpha_{12})$ and $i_\mathrm{mut}$ is the mutual inclination of the hot Jupiter and the companion (based on Equation~64 from \citealt{Su&Lai2022}).\footnote{For WASP-12\,b, this complementary constraint was not needed. Angular-momentum considerations and the RV data ruled out the full parameter space.}

Overall, the requirements on a companion capable of sustaining a high-obliquity Cassini state and a high dissipation rate are severe enough to make the scenario seem unlikely in general.

\subsection{Summary}
\label{sec:conclusion}

The obliquity-tide hypothesis is appealing because the energetics are favorable and it can operate even when the orbit is nearly circular. Although it may solve the energy problem, it creates an angular-momentum problem. The companion must not only set the nodal precession frequency needed for the spin-orbit resonance, it must also act as a flywheel, absorbing the angular momentum that is lost as the hot Jupiter spirals inward. At the same time, it must not be too massive or close to the hot Jupiter to provoke dynamical instability. Together, these requirements made the hypothetical companion easier to rule out with current RV data than was anticipated by \cite{Millholland&Laughlin2018}. We must therefore look elsewhere to explain the rapid orbital decay of WASP-12\,b.

\begin{acknowledgements}

We thank Sarah Millholland for stimulating discussions that inspired this work. We also thank Dan Fabrycky, Jeremy Goodman, and the anonymous referee for valuable comments. C.L. acknowledges support from a Natural Sciences and Engineering Research Council of Canada (NSERC) Postgraduate Scholarship.

 We are pleased to acknowledge that the work reported in this paper was substantially performed using the Princeton Research Computing resources at Princeton University, which is a consortium of groups led by the Princeton Institute for Computational Science and Engineering (PICSciE) and the Office of Information Technology's Research Computing.

\end{acknowledgements}

\appendix

\section{RV dataset}
\label{sec:RV_dataset}

The RV measurements compiled for our analysis are included in Table~\ref{table:RV_data}. All measurements come from the literature, except four NEID RVs that were collected after the publication of \citet{WinnStefansson2025}.

\begin{deluxetable*}{ccccc}
\tablecaption{WASP-12's RV Measurements\label{table:RV_data}}
\tablehead{
  \colhead{Time} & \colhead{RV} & \colhead{RV uncertainty} &
  \colhead{Instrument} & \colhead{Source} \\
  \colhead{[BJD]} & \colhead{[m/s]} & \colhead{[m/s]} &
  \colhead{} & \colhead{}
}
\startdata
$2454509.38633$ & $18923.1$ & $8.8$ & SOPHIE & \citet{Hebb2009} \\
$2454849.42764$ & $19202.8$ & $11.2$ & SOPHIE & \citet{Husnoo2011} \\
$2455188.03054$ & $-168.8$ & $3.3$ & HIRES & \citet{Yee2020} \\
$2456244.73506$ & $18922.1$ & $5.7$ & HARPS-N & \citet{Bonomo2017} \\
$2456294.63701$ & $18982.9$ & $2.8$ & HARPS-N & \citet{Maciejewski2020} \\
$2459473.93624$ & $18898.9$ & $8.0$ & NEID & \citet{WinnStefansson2025} \\
$2460915.99340$ & $18834.8$ & $5.4$ & NEID & This work \\
$\cdots$ & $\cdots$ & $\cdots$ & $\cdots$ & $\cdots$
\enddata
\tablecomments{The full table is available online in the published version of this paper. Only the first data point from each source is included here.}
\end{deluxetable*}

\bibliography{refs}{}
\bibliographystyle{aasjournalv7}

\end{document}